\documentclass[prd,twocolumn,floatfix,preprintnumbers,letterpaper,nofootinbib]{revtex4}
\usepackage{graphicx}
\usepackage{amsmath,amssymb}
\input{epsf}
\usepackage{epsf}
\usepackage{footnote}
\usepackage{natbib}

\begin{document}

\def\be{\begin{equation}}
\def\ee{\end{equation}}

\title{Apocalypse When?  Solar System Constraints on an Imminent Big Rip}
\author{Robert J. Scherrer$^{1}$ and Oem Trivedi$^1$}
\affiliation{$^1$Department of Physics \& Astronomy, Vanderbilt University,
Nashville, TN~~37235}
\date{\today}

\begin{abstract}
Phantom dark energy models with an equation of state parameter $w < -1$ lead generically to a future big rip singularity, in which the dark energy density becomes infinite in a finite time.  Current limits on dark energy constrain $w$ to be close to $-1$, and if $w$ is assumed constant, then a future big rip cannot occur in less than the order of a Hubble time in the future.  However, many models allow $w$ to decrease rapidly with time.  In that case, or if one assumes an additional phantom component with current energy density far below the dark energy density and $w << -1$, it is possible to achieve an imminent big rip, which we define to be a future singularity occuring in much less than the Hubble time.  Such a possibility cannot be constrained by any cosmological measurements, as
these are all based on light emitted billions of years in the past.  Indeed, it is not possible, on the basis of cosmological observations, to rule out a future big rip tomorrow.  However, solar system dynamics are sensitive to the behavior of phantom dark energy on timescales of decades
rather than billions of years.  Using solar system measurements, we are able to derive limits
on the timescale for a future big rip independent of the dynamics of the phantom component.  We obtain $t_{rip} - t_0 > 30$ years.  While admittedly a poor limit, these results are likely to be improved by future more precise measurements of solar system dynamics.  Our results also show that evidence for an imminent big rip would show up first in solar system data, rather than in any cosmological observation.
\end{abstract}

\maketitle

Dark energy is characterized by the
equation-of-state parameter $w$, defined by
\begin{equation}
w = p_{DE}/\rho_{DE},
\end{equation}
where $\rho_{DE}$ and $p_{DE}$ are the density
and pressure, respectively, of the dark energy.
It has long been known that observational data are consistent with ``phantom" dark energy,
for which $w < -1$ \cite{Caldwell}.
Such ``phantom" dark energy
models have several peculiar properties.
The density of the dark energy {\it increases} with
increasing scale factor, and
the phantom energy density can become infinite
at a finite time $t$, a condition known as the ``big rip" \cite{Caldwell,rip,Taka,rip2,scherrerrip}.

Models of this kind can arise in a variety of contexts, including scalar field models with a negative kinetic term, or barotropic models, in which the pressure is a specified function of the density.  Phantom models necessarily violate the weak energy condition and have
certain well-known problems \cite{Carroll,Cline,Hsu1,Hsu2}.  Nonetheless, they remain of interest in cosmology; for instance, the recent DESI results can be well fit by assuming a transient epoch with $w < -1$ \cite{DESI1,DESI2,DESI3}.

If the currently-observed dark energy has an equation of state parameter $w$, and $w$ is
assumed to be constant, then it is easy to show that a future big rip singularity will occur at a time
\cite{scherrerrip}
\begin{equation}
t_{rip} = \left(\frac{w}{1+w}\right)t_m,
\end{equation}
where $t_m$ is the time at which the matter and dark energy densities were equal.  Unless $w << -1$,
$t_{rip}$ is of order the Hubble time, which is safely billions of years in the future.

Here we consider a different possibility, that of an imminent big rip, which we define to be a future singularity in a time much less than the Hubble time.  One possibility for an imminent big rip is a model in which the currently-observed dark energy has $w$ slightly less than $-1$, but which transitions on a very short timescale to $w << -1$.  
As an example, consider the modified Chaplygin gas, with an equation of state given by
\begin{equation}
p = - \frac{A}{\rho^\alpha}
\end{equation}
where $A$ and $\alpha$ are constants in the model.
Defining $A_s = A/\rho_0^{1+\alpha}$, where $\rho_0$ is the present-day dark energy density, and taking the scale
factor $a$ to be equal to 1 at present, the density evolution in this model is \cite{MCG}
\begin{equation}
\rho = \rho_0 \left[A_s + (1-A_s) a^{-3(1+ \alpha)}\right]^{1/(1+\alpha)},
\end{equation}
with a corresponding evolution for $w$:
\begin{equation}
w = - \frac{A_s}{A_s + (1-A_s)a^{-3(1+\alpha)}}.
\end{equation}
As noted in Ref. \cite{MCG}, models with $A_s > 1$ and $\alpha < -1$ evolve initially close to
a cosmological constant, with $w \approx -1$, and then at late times they undergo phantom behavior with $w \rightarrow -\infty$.
In these models, a big rip occurs when $A_s + (1-A_s) a^{-3(1+ \alpha)} = 0$.  In particular, consider a model
for which $A_S = 1 + \delta$, with $\delta << 1$.  This gives a phantom model with $w = -(1+\delta)$ at present, nearly indistinguishable from a cosmological constant.  However, it will transition sharply and give a future big rip
at a scale factor corresponding to $a^{3(1+\alpha)} = \delta$.  By choosing $\alpha \ll -1$, we can achieve a big rip for $a$ arbitrarily close to 1, i.e. a big rip on a timescale much smaller than the Hubble time.

Behavior of this kind can also be produced by a scalar field with a negative kinetic term and an appropriately pathological potential $V(\phi)$.  For example, consider a potential $V(\phi)$ that is nearly flat for $\phi < \phi_0$ and increases sharply for $\phi > \phi_0$, where $\phi_0$ is the present-day value of $\phi$.  Such a scalar
field can mimic a cosmological constant (with $w$ just below $-1$) all of the way up to the present, but a sufficiently steep potential for $\phi > \phi_0$ can lead to an arbitrarily imminent big rip.  (See also the discussion in
Ref. \cite{Astashenok}).

Models of this kind are not particularly compelling, and we do not consider them likely, but surprisingly, they cannot be definitively ruled out.  As noted by Andrei, Ijjas, and Steinhardt \cite{Andrei}, any change to the expansion law on a timescale much shorter than the Hubble time cannot be detected with cosmological observations such as the cosmic microwave background or distant supernovae, since the corresponding light was emitted long before the change in the expansion.  Even the Sandage-Loeb effect \cite{Sandage,Loeb1}, which is the closest thing we have to a ``real-time" cosmological observation, relies on the light from high-redshift quasars.  Thus, it is quite possible that $\rho_{DE}$ is currently much larger than cosmological estimates, and $w$ could be much further below $-1$ than implied by cosmological observations.
However, there is one set of observational tests which are sensitive to a sharply increasing dark energy density on a timescale much less than the Hubble time:  solar system orbital dynamics, which are sensitive
to the values of $\rho_{DE}$ and $w$ on timescales of decades rather than billions of years.

To make further progress, we need to assume something about the evolution of $w$ and $\rho_{DE}$.
We will assume that the phantom dark energy corresponds to the presently-observed dark energy.  However, we also
assume that the dark energy has undergone a sharp transition recently to a value of $w << -1$ on a timescale much less than the Hubble time, so that the true present value of $\rho_{DE}$, which we will denote $\rho_0$, could be much larger than the cosmologically-inferred dark energy density, while the true present value of $w$ could be much more negative than the cosmologically-inferred value.  (Note that we could also consider models in which the phantom component has nothing to do with the currently-observed dark energy.  In this case the phantom component is currently subdominant, but with $w \ll -1$, so that the phantom energy
density is rapidly increasing at present and will soon dominate the expansion.  We will see that these models are more difficult to constrain than models in which the observed dark energy transitions rapidly to phantom behavior).

For simplicity, we will assume constant $w$, so that
\begin{equation}
\label{rhode}
\rho_{DE} = \rho_0 a^{-3(1+w)},
\end{equation}
where $a = 1$ corresponds to the present.
This should be a reasonable approximation over sufficiently short timescales, and in any event, our results are easily generalized to other cases of interest.  Then the time variation of $\rho_{DE}$ is given by
\begin{equation}
\label{drhodt}
\dot \rho_{DE} = -3 (1+w) \rho_{DE} \sqrt{\frac{8}{3} \pi G \rho_{T}}, 
\end{equation}
where $\rho_T$ is the total density.  If the phantom dark energy density dominates on a timescale much shorter than the Hubble time, then we can take $\rho_T = \rho_{DE}$ and integrate $dt/d \rho_{DE}$ to find the time
$t_{rip}$ at which a future big rip occurs relative to the present time $t_0$:
\begin{equation}
\label{trip}
t_{rip} - t_0 = \frac{1}{|1+w|\sqrt{6 \pi G \rho_0}}
\end{equation}
Thus, a lower bound on $t_{rip} - t_0$ requires an upper bound on $|1+w| \sqrt{\rho_0}$.

As noted above, cosmological observations cannot provide any constraints on such models, but solar system dynamics can provide limits.
For a mass $m$ orbiting a central mass $M$ at a distance $r$, the total force is
\begin{equation}
\label{force}
F = - \frac{GMm}{r^2} - \frac{4 \pi}{3} Gmr \rho_{DE}(1+3w),
\end{equation}
where $3 w\rho_{DE}$ gives the effect of the pressure from the dark energy.
Jetzer and Sereno used the
orbits of the Earth and Mars to give an upper limit on the magnitude of a cosmological constant, deriving a limit of \cite{JetzerSereno}
\begin{equation}
\Lambda < 1 \times 10^{-36} {\rm km}^{-2},
\end{equation}
corresponding to an upper limit on the (assumed constant) dark energy density of
\begin{equation}
\label{rholambda}
\rho_{0} < 5 \times 10^{-20} {\rm g}~ {\rm cm}^{-3}.
\end{equation}
As noted in Ref. \cite{JetzerSereno}, this limit is $10^{10}$ times larger than the cosmological upper bound.
If we assume instead a time-varying dark energy component with $w \ne -1$, then using Eq. (\ref{force}) the
Jetzer-Sereno bound
becomes an upper limit on $\rho_{DE} (1+3w)$, namely
\begin{equation}
\label{K0}
\rho_{0} |1+ 3w| < 1 \times 10^{-19}  {\rm g}~ {\rm cm}^{-3}.
\end{equation}

A second limit on $\rho_0$ and $w$ can be derived from constraints on the time variation of $\ddot a/a$.
To derive such a bound, we use the Friedman equation in the form
\begin{equation}
\frac{\ddot{a}}{a} = - \frac{4 \pi G}{3} \rho_{DE} (1+3w).
\end{equation}
Assuming constant $w$, the time derivative is
\begin{equation}
\label {dadt}
\frac{d}{dt}\left(\frac{\ddot{a}}{a}\right) = - \frac{4 \pi G}{3} \dot\rho_{DE} (1+3w).
\end{equation}
Using Eq. (\ref{drhodt}) with $\rho_{DE} = \rho_T =  \rho_0$ gives
\begin{equation}
\label{K1}
\frac{d}{dt}\left(\frac{\ddot{a}}{a}\right) = \sqrt{\frac{128}{3} \pi^3 G^3 \rho_{0}^3} (1+3w)(1+w).
\end{equation}
Iorio \cite{Iorio} 
explored solar system limits on the time derivative of $\ddot a/a$.  His best limit comes from the orbit of Saturn, for which \cite{Iorio}
\begin{equation}
\label{Iolimit}
\frac{d}{dt}\left(\frac{\ddot{a}}{a}\right) < 2 \times 10^{-13} {\rm yr}^{-3}.
\end{equation}
Other solar-system limits on the time derivative of $\ddot a/a$ can be inferred from upper limits on the time variation
of $G$ 
derived from ranging measurements to
Mars and the Moon \cite{Will}.
These limits are, for Mars \cite{Konopliv}
\begin{equation}
\label{mars}
\dot G/G = 0.1 \pm 1.6 \times 10^{-13} ~{\rm yr}^{-1},
\end{equation} 
and for the Moon \cite{Williams1}
\begin{equation}
\label{moon}
\dot G/G = 4 \pm 9 \times 10^{-13}~{\rm yr}^{-1}.
\end{equation}
These limits translate into a limit on the time variation of the gravitational force, as $\dot G/G = \dot F/F$.  However, a phantom component would also produce a (very weak) time varying force, which could be misinterpreted as a time-varying $G$.

Assuming that $G$ and $w$ are constant, but $\rho_{DE}$ is rapidly varying, Eq. (\ref{force}) gives
\begin{equation}
\frac{\dot F}{F} = \frac{4\pi}{3} \frac{\dot \rho (1+3w)}{M/r^3}.
\end{equation}
Then using Eq. (\ref{dadt}), we have
\begin{eqnarray}
\label{marsmoonlimits}
\frac{d}{dt}\left(\frac{\ddot{a}}{a}\right) &=& - \frac{\dot F}{F} \frac{GM}{r^3}, \\
&=& - \frac{\dot F}{F} (2 \pi/T)^2,
\end{eqnarray}
where $T$ is the orbital period of the body in question.
We take $\dot F/F = \dot G/G$ and note that phantom dark energy would have the same effect as a decreasing $G$,
so we need to use the lower bounds in Eqs. (\ref{mars}) and (\ref{moon}).
Clearly, the Martian limit is the stronger one, and we obtain
\begin{equation}
\frac{d}{dt}\left(\frac{\ddot{a}}{a}\right) < 2 \times 10^{-12} {\rm yr}^{-3}.
\end{equation}
Thus, Iorio's limit (Eq. \ref{Iolimit}) is the strongest, by an order of magnitude.

Inserting this bound into Eq. (\ref{K1}) and taking the limit where $w << -1$, we obtain
\begin{equation}
\label{limit1}
\rho_0 |w|^{4/3} < 2 \times 10^{-18} {\rm g}~ {\rm cm}^{-3}.
\end{equation}
Taking the same limit in Eq. (\ref{K0}) gives       
\begin{equation}
\label{limit2}
\rho_0 |w| < 3 \times 10^{-20} {\rm g}~ {\rm cm}^{-3}.
\end{equation}
Rather surprisingly, even in combination these two limits do not provide an upper limit on
$|1+w| \sqrt{\rho_0}$ in Eq. (\ref{trip}).  The reason is that both bounds can be satisfied, and 
$|1+w| \sqrt{\rho_0}$ can be made arbitrarily large, but taking $\rho_0 \rightarrow 0$ and $w \rightarrow - \infty$.
Thus, in models in which $\rho_0$ can be arbitrarily small and $w$ can be arbitrarily negative,
no solar system limit is possible.  However, we can place a limit using our assumption that
the phantom dark energy is the cosmologically-observed dark energy.  Since this energy density can only increase
with time, we can place a lower bound on $\rho_0$, namely that it must be larger than the cosmologically-measured
dark energy density,
\begin{equation}
\label{rholimit}
\rho_0 > 6 \times 10^{-30} {\rm g}~{\rm cm}^{-3}.
\end{equation}
For this lower bound on $\rho_0$, Eq. (\ref{limit1}) gives a better bound on $w$ than does Eq. (\ref{limit2}),
namely
\begin{equation}
|w| <  4 \times 10^8.
\end{equation}
When this lower limit on $\rho_0$ and upper limit on $|w|$ are saturated, Eq. (\ref{trip}) gives the lower bound
on $t_{rip}$ from solar system data.  This limit is
\begin{equation}
\label{finallimit}
t_{rip} - t_0 > 30 ~{\rm yr}.
\end{equation}
While this bound on $t_{rip}$ may seem surprisingly weak, what is actually surprising is that one can place any limits at all on an imminent big rip.  The big rip cannot happen tomorrow, at least not for models in which the phantom dark energy is identified with the currently observed dark energy.

We have assumed here that the phantom field driving a future big rip couples only gravitationally.  If it had even a tiny coupling to electromagnetic fields, then any imminent big rip would show up as a change in the fine structure constant or in the electron-proton mass ratio, both of which are strongly constrained by laboratory measurements.  However, in the absence of such a coupling, Eq. (\ref{finallimit}) is the best that we can do,
although future more precise measurements of solar system dynamics will improve this limit.

These results suggest than an imminent big rip would first manifest itself in solar system dynamics.  This is rather counterintuitive, since a big rip proceeds from the outside in, i.e., from the largest scales (which are the most weakly bound), down to smaller scales \cite{rip}.  However, as we (and Ref. \cite{Andrei}) have emphasized, cosmological bounds measure the state of the universe billions of years ago and are insensitive to changes to the evolution of the universe on timescales much smaller than the Hubble time.  For an imminent big rip, the best we can say is that it will not happen soon.

\end{document}